\documentclass[useAMS,usenatbib]{mn2e}
\usepackage{graphicx}
\newcommand{\new}[1]{{#1}}
\newcommand{\newII}[1]{{#1}}

\def\ltsim{\mathrel{\hbox{\rlap{\hbox{\lower4pt\hbox{$\sim$}}}\hbox{$<$}}}}

\title[Origin of the profile of WR\,2 spectral lines]{Investigating the origin of the spectral line profiles of the Hot Wolf-Rayet Star WR\,2}

\author[A.-N. Chen\'e et al.]
{\parbox{\textwidth}{A.-N. Chen\'e$^{1,2,3}\thanks{\texttt{Based on observations obtained at the Canada-France-Hawaii Telescope (CFHT) which is operated by the National Research Council of Canada, the Institut National des Sciences de l'Univers of the Centre National de la Recherche Scientifique of France, and the University of Hawaii. Also based on observations obtained at the Gemini Observatory, which is operated by the Association of Universities for Research in Astronomy, Inc., under a cooperative agreement with the NSF on behalf of the Gemini partnership: the National Science Foundation (United States), the National Research Council (Canada), CONICYT (Chile), the Australian Research  Council (Australia), Minist\'erio da Ci\^encia, Tecnologia e Inova\c c\~ao (Brazil) and Ministerio de Ciencia, Tecnolog\'ia e Innovaci\'on Productiva (Argentina).}}$\thanks{\texttt{E-mail: andrenicolas.chene@gmail.com (ANC)}},
N. St-Louis$^{4}$,
A.~F.~J. Moffat$^{4}$,
O. Schnurr$^{5}$,
P.~A. Crowther$^{6}$,
G.~A. Wade$^{7}$,
N.~D. Richardson$^{8}$,
C. Baranec$^{9}$,
C.~A. Ziegler$^{10}$,
N.~M. Law$^{10}$,
R. Riddle$^{11}$,
G.~A. Rate$^{6}$,
\'E.~Artigau$^{4}$,
E. Alecian$^{12}$  and
BinaMIcS collaboration}\vspace{0.4cm}\\
\parbox{\textwidth}{$^{1}$Gemini Observatory, Northern Operations Center, 670 North A'ohoku Place, Hilo, HI 96720, USA\\
$^{2}$Departamento de F\'isica y Astronom\'ia, Universidad de Valpara\'iso, Av. Gran Breta\~na 1111, Playa Ancha, Casilla 5030, Chile\\
$^{3}$Departamento de Astronom\'ia, Universidad de Concepci\'on, Casilla 160-C, Chile\\
$^{4}$D\'epartement de Physique, Universit\'e de Montr\'eal, C. P. 6128, succ. centre-ville, Montr\'eal (Qc) H3C 3J7, and\\
 Centre de Recherche en Astrophysique du Qu\'ebec, Canada\\
$^{5}$Leibniz-Institut f\"ur Astrophysik Potsdam (AIP), An der Sternwarte 16, 14482 Potsdam, Germany\\
$^{6}$Department of Physics and Astronomy, University of Sheffield, Sheffield S3 7RH, UK\\
$^{7}$Department of Physics, Royal Military College of Canada, PO Box 17000, Stn Forces, Kingston, K7K 7B4, Canada\\
$^{8}$Ritter Observatory, Department of Physics and Astronomy, The University of Toledo, Toledo, OH 43606-3390, USA
$^{9}$Institute for Astronomy, University of Hawai`i at M\={a}noa, Hilo, HI 96720-2700, USA\\
$^{10}$Dunlap Institute for Astronomy and Astrophysics, University of Toronto, Ontario M5S 3H4, Canada
$^{11}$Division of Physics, Mathematics, and Astronomy, California Institute of Technology, Pasadena, CA 91125, USA\\
$^{12}$Universit\'e Grenoble Alpes, CNRS, IPAG, F-38000 Grenoble, France}}

\begin{document}

\date{Accepted  XXX YY. Received xxx YY}

\pagerange{\pageref{firstpage}--\pageref{lastpage}} \pubyear{2014}

\maketitle

\label{firstpage}

\begin{abstract}
The hot WN star WR\,2 (HD\,6327) has been claimed to have many singular characteristics. To explain its unusually rounded and relatively weak emission line profiles, it has been proposed that WR\,2 is rotating close to break-up with a magnetically confined wind. Alternatively, the line profiles could be explained by the dilution of WR\,2's spectrum by that of a companion. In this paper, we present a study of WR\,2 using near-infrared AO imaging and optical spectroscopy and polarimetry.
Our spectra reveal the presence of weak photospheric absorption lines from a $\sim$B\,2.5-4V companion, which however contributes only $\sim$5--10\% to the total light, suggesting that the companion is a background object. Therefore, its flux cannot be causing any significant dilution of the WR star's emission lines.
The absence of intrinsic linear continuum polarization from WR\,2 does not support the proposed fast rotation. Our Stokes V spectrum was not of sufficient quality to test the presence of a moderately strong organized magnetic field but our new modelling indicates that to confine the wind the putative magnetic field must be significantly stronger than was previously suggested sufficiently strong as to make its presence implausible.
\end{abstract}


\section{Introduction}\label{intro}

Classical Wolf-Rayet (WR) stars are the chemically evolved descendants of O stars, possessing a unique spectroscopic signature of strong, broad emission lines, arising within their fast, dense winds \citep{C07}. Their two main subclasses, WN and WC, reflect the products of core hydrogen and helium burning, respectively. The former is commonly subdivided into early (WN2--6) and late (WN7--11) types, depending upon the degree of ionization of helium and nitrogen \citep{SSM96}. Early-type WN stars in the Milky Way exhibit either strong, broad Gaussian line profiles (e.g. WR6, WN4b), or weak, narrow triangular profiles (e.g. WR152, WN3(h)), with one notable exception~: WR\,2 (WN2b), the subject of this study.

WR\,2 (HD~6327) is the only known WN2 star in our Galaxy. Its spectrum displays relatively weak ``bowler-hat''-shaped broad emission lines, in stark contrast with the more normal strong (Gaussian) or weak (triangular) profiles of other early-type WN stars. \new{Considering that WR\,2 has an extreme combination of a small radius, fairly low mass-loss rate and high temperature \citep{Ha06}, a certain line dilution due to the Baldwin effect \citep{vG01} is expected. However, as we will present below, this does not seem to be sufficient to explain the line profiles.} 

Two scenarios have been proposed to explain this unusual characteristic. The first involves a companion star, where intrinsically strong Gaussian lines appear weak owing to significant line dilution from a companion \citep{C93}. However, convincing evidence of such a companion has not been reported in the literature to date. The second scenario involves the intrinsic Gaussian/triangular lines being modified by rapid rotation. The Potsdam group have analyzed WR\,2 using the latest version of their model atmosphere code \citep{Ha06}, and remarkably the model spectrum failed to reproduce its line profiles, unless it was folded with a rotation profile near break-up velocity, with an equatorial rotation speed of 1900 km s$^{-1}$.\new{ Of course, as mentioned by the authors themselves, it is a purely mathematical solution, as a mere solid rotation is not physical and many complex interactions are expected when such a fast rotation speed occurs.} More recently, \citet{Sh14} have developed a more detailed atmospheric model which proposes a much slower\new{, and realistic} surface rotation speed ($\sim$ 200 km\,s$^{-1}$), but a very strong magnetic field of $\sim$40\,kG at the surface to force rigid rotation of the wind up to a radius $\sim$10\,$R_\ast$.

Four more ``round-lined'' WR stars are known in the LMC, BAT99 7, 51, 88 and 94 \citep{Ru08,Sh14}, but this type of line profile still remains a very rare oddity. As WR\,2 has an apparent magnitude that is at least 4 magnitudes brighter than the more distant LMC stars, it is a perfect proxy to study the phenomenon.

If WR\,2's claimed rapid rotation were confirmed, it would make it a prime long-soft gamma-ray burst (LGRB) progenitor candidate. LGRBs are currently best explained by the ``collapsar'' model \citep{Wo93,MF99,MF01}, involving a rapidly rotating, fully hydrogen-depleted WR star undergoing core collapse. Classical WR stars are the helium-burning cores of evolved massive stars in their last visible evolutionary phase before they explode as type Ib/c core-collapse supernovae (ccSNe). Direct observational evidence of a type-Ic--LGRB connection \citep{Ke08} lends considerable weight to the collapsar model, and might imply that LGRB progenitors are merely the rapidly rotating tail in the distribution of type Ic progenitors. The mere presence of rapid rotation can in principle be revealed quite easily by linear spectropolarimetry \citep{Ha00}: the rotation causes axisymmetric distortions of stellar winds, which in turns causes polarization from scattered radiation, and since the continuum radiation is formed at smaller radii than the line emission, the net effect is an apparent decrease of polarization in the lines.

On the other hand, if the proposed strong magnetic field of WR\,2 was confirmed, it would make it a prime magnetar progenitor candidate. Magnetars are neutron stars with surface magnetic fields as strong as $10^{14}$--$10^{15}$\,G \citep[e.g.][]{Fe06}. The current explanation for the origin of a neutron star's magnetic field is the conservation of the star's magnetic flux during core collapse; magnetars could then be descendants of the most strongly magnetized WR stars \citep{Ga05}. The shape of the WR wind would also be affected by its confinement by the strong magnetic field \citep{Ow04,To05b,ud13}, and its asymmetry could be detected in linear spectropolarimetry. 

This study presents our investigation of the origin of the rounded emission line profiles of WR\,2. In Section\,\ref{Comp} we search for any manifestation of a companion star and investigate how its presence can lead to round emission lines. We search for evidence of rapid stellar rotation in Section\,\ref{Rot}, where we present new linear broad-band polarimetry as well as spectropolarimetry to assess the symmetry of WR\,2's wind. We look for evidence of the presence of a strong magnetic field in Section\,\ref{magfld}, where we discuss our attempt to detect a global magnetic field in WR\,2 using the Zeeman effect (Section\,\ref{specpolV}). Finally, all our results are summarized in Section\,\ref{Disc}, where we present what we have learned from these observations about the true nature of WR\,2.

\begin{table*}
\caption{Spectroscopic Data}
\begin{tabular}{lccrr|lccrr}
\hline
Telescope & Dates & $\lambda$ Coverage & $\Delta\lambda$ & S/N & Telescope & Dates & $\lambda$ Coverage & $\Delta\lambda$ & S/N\\
Instrument  &   & \AA & (3 pixels)& & Instrument  &   & \AA & (3 pixels)\\
\hline
INT/IDS & 1991 Sep 19--22 & 3800--7290 & 2--2.7\,\AA &  30 & DAO$\rm{^a}$ & 2008 Aug 15 & 4300--5090 & 2.3\,\AA & 340\\
OMM$\rm{^a}$ & 2001 Aug 12, 25 & 4490--4940 &  0.7\,\AA & 190 & OMM & 2008 Dec 21, 25 & 3680--5380 &  1.9\,\AA & 80\\
OMM$\rm{^b}$ & 2002 Jul 21, 25; & 4500--5200 &  1.6\,\AA & 100 & CFHT$\rm{^c}$ & 2010 Aug 03 & 3700--10480 &  0.06\,\AA & 100\\
  & 2002 Oct 19--23 &  &  & & Gemini/ & 2012 Jul 11--23  & 3970--5435 &  1.5\,\AA & 500\\
WHT/ISIS & 2002 Nov 19 & 4245--5130 & 0.6\,\AA & 200 & \,\,\,\,\, GMOS$\rm{^c}$ & Aug 31 &  &  & \\
\hline
\multicolumn{5}{l}{\hspace{0.5cm} $\rm{^a}$ published by \citet{Ch08a}}\\
\multicolumn{5}{l}{\hspace{0.5cm} $\rm{^b}$ published by \citet{St09}}\\
\multicolumn{5}{l}{\hspace{0.5cm} $\rm{^c}$ this study}\\
\end{tabular}\label{spdata}
\end{table*}

\section[]{Search for a companion star}\label{Comp}

Before testing any extreme stellar parameter of WR\,2, we first verify the possibility that its lines are diluted by the flux of a companion. 

\subsection{Observations}
\subsubsection{Spectroscopy}
This study is based on a series of published and unpublished, archival and new spectra from different observatories. This includes spectra from the 1.6m telescope of the Observatoire du Mont M\'egantic (OMM), the 4.2m William Herschel Telescope (WHT) and the 1.85m telescope of the Dominion Astrophysical Observatory (DAO) obtained over nearly ten years. We add to that a high-resolution spectrum obtained with the ESPaDOnS spectropolarimeter at the 3.6m Canada-France-Hawaii Telescope (CFHT) that we present in more detail in Section\,\ref{specpol}, and high S/N spectra obtained using GMOS at the 8.2m Gemini-North telescope (originally obtained to search for line profile variations). Finally, we also include 2.5m Isaac Newton Telescope (INT) spectrophotometry which is used for modelling in Section\,\ref{flrat}. When series of spectra were obtained, all were combined into nightly averages to optimize the S/N. The details of the spectroscopic observations are summarized in Table\,\ref{spdata}.

\subsubsection{Direct imaging}
In an effort to spatially isolate a putative companion of WR\,2, we first acquired visible-light adaptive optics imaging of WR\,2 using the Robo-AO system \citep{Ba14} on the 1.5m Telescope at Palomar Observatory on 2014 November 9. The observation in the Sloan i band comprises a sequence of full-frame-transfer detector readouts at the maximum rate of 8.6 Hz for a total of 120 s of integration time. The individual images were then combined using post-facto shift-and-add processing using WR\,2 as the tip-tilt star with 100\% frame selection \citep{La14}.

We also obtained time for a Poor Weather program at Gemini-North to observe a diffraction-limited image with NIRI+Altair in the $K$ band. Those data required basic data reduction processes using the Gemini {\sc iraf}\footnote{{\sc iraf} is distributed by the National Optical Astronomy Observatories (NOAO), which is operated by the Association of Universities for Research in Astronomy, Inc. (AURA) under cooperative agreement with the National Science Foundation (NSF).} packages for NIRI.

\subsection[]{Search for a companion spectrum}

\begin{figure*}
\includegraphics[width=18cm]{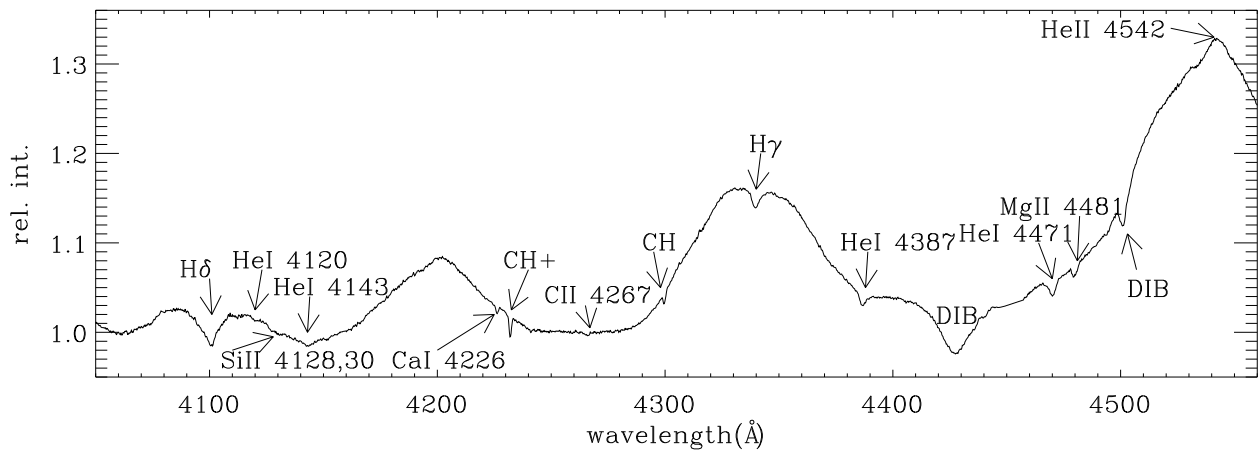}
 \caption{High S/N average GMOS spectrum obtained between 2012 July 11 and August 31. We show here only the wavelength range between 4050\AA\, and 4560\AA, where most of the absorption lines are visible. We label the IS lines as well as the H, He\,{\sc i}, Si\,{\sc ii}, C\,{\sc ii} and Mg\,{\sc ii} lines seen superimposed on the WR emission lines (unlabeled). Note the absence of a narrow absorption line at He\,{\sc ii}$\lambda$4542.}
  \label{figcomp}
\end{figure*}

An inspection of the WR\,2 spectrum reveals clear narrow absorption lines mostly associated with H and He\,{\sc i} in all our spectra. To identify the spectral lines, we have averaged all GMOS spectra to reach a S/N$\sim$1300 per resolution element in the continuum. The result is presented for the region 4050 -- 4570\,\AA\, in Fig.\,\ref{figcomp}, where the absorption features from a companion and those formed in the interstellar medium (ISM) are marked and identified. A few more absorption lines are visible in the rest of the spectra, and are mostly associated to H or are interstellar (IS) lines. The equivalent width ratio of the He\,{\sc i}$\lambda$4471 to the Mg\,{\sc ii}$\lambda$4481 lines is 1.95, which corresponds to a spectral type of B\,2.5--4 (a B\,2.5 type will be used from now on for convenience) and the ratio of the S\,{\sc iii}$\lambda$4552 (barely visible in our spectrum) to the He\,{\sc i}$\lambda$4387 line suggests a luminosity class of V \citep{Wa90}. The very faint Si\,{\sc ii}$\lambda$4128,30 and C\,{\sc ii}$\lambda$4267 lines can barely be distinguished.

The closest known star to WR\,2 TYC4017-1829-2, has a V~=~14.81\,mag (compared to V~=~10.99\,mag for WR2) and is separated by $13.8^{\prime\prime}$. This is typically 5 times the size of the seeing disk at the smaller telescopes used for this study, and more than 10 times the size of the seeing disk at the Maunakea Observatories. Hence conclude that there is no significant contamination of WR\,2's spectrum by this star, and the absorption lines likely come from a companion to WR\,2.

\subsection[]{Search for radial velocity variations} \label{RV}

\begin{figure*}
\includegraphics[width=18cm]{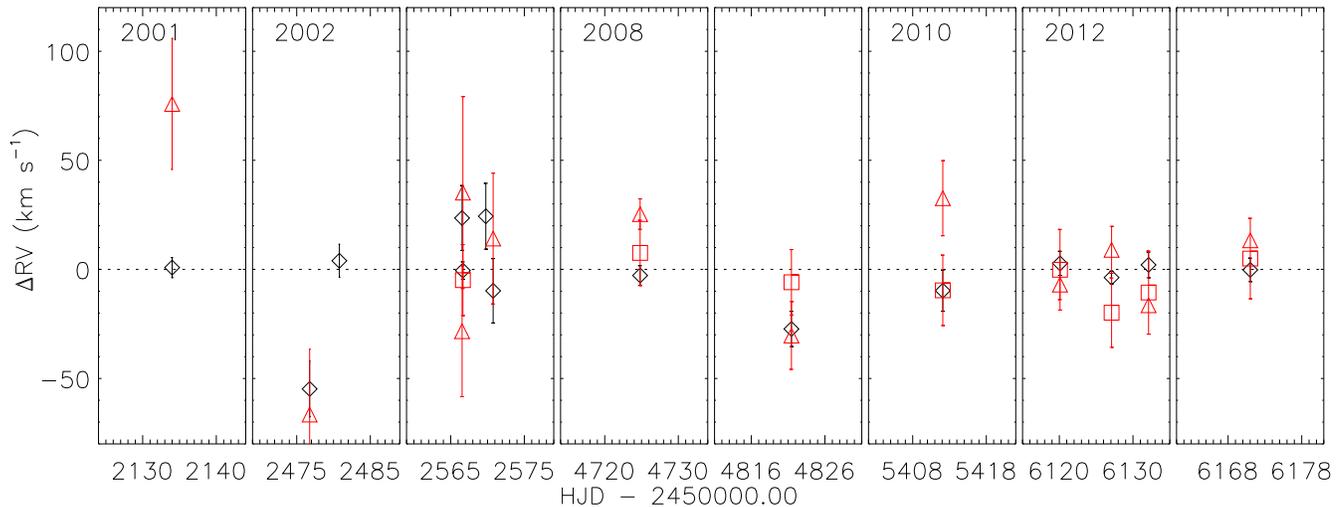}
 \caption{Radial velocity (RV) variations of WR\,2 measured using all our spectra observed between 2001 and 2012. The emission line RV shifts (black diamonds) were obtained by cross-correlating the spectra with the GMOS spectrum of 2012 August 31 that served as a reference. The absorption line RVs were measured by fitting their profiles using a Voigt function. We assume that the lines of a given element will all move together, hence, we combine the RVs for all H (red triangles) and He lines (red squares). The $\pm$1-$\sigma$ error bars associated with each measurement are superposed on the data points.}
  \label{rv}
\end{figure*}

In order to determine if the B companion is bound to WR\,2 or is an accidentally aligned star at a different distance, we first searched for radial velocity (RV) changes in both the emission and the newly discovered absorption lines. We used the many IS lines across the spectrum as references for RV, which allows one to correct for zero-point shifts in spectra coming from various instruments. The search for RV variations of the WR star was done using cross-correlation of the spectra in the region from 4300\,\AA\, to 5000\,\AA, with the GMOS spectrum from the night of 2012 August 31 as reference. The signal of the cross-correlation profile is dominated by the WR emission lines. \new{The error on the values is given by the width of the cross-correlation profile, which typically varies between 10 and 30 km\,s$^{-1}$. Those errors remain small regardless of the broadness of the WR lines due to the wide wavelength coverage of the spectra.} The result is presented in Fig.\,\ref{rv}, using black diamond symbols. The amplitude of the measured RV variation is very small, and when it is compared to the 1-$\sigma$ error bars, we see that the variations rarely exceed the 1- or 2-$\sigma$ level. The RV of the absorption line star was obtained by fitting the absorption lines with a Voigt profile. For clarity, we assume that the lines of a given element move together, hence we plotted only an average value for the H (red triangles) and He (red squares) lines. The RV values for the H and He lines were corrected by subtracting their time-median value, so they can be compared with the WR RV values centered on 0 km\,s$^{-1}$. \new{The error on the values depends highly on the spectra signal-to noise ratio, since the absorption lines are shallow, and in fairly small number per specie.} Once again, when compared with the error bars, the variations barely exceed 1-$\sigma$ (except in few rare cases when the emission and absorption lines seem to have fortuitously moved together). We therefore conclude that we do not detect any significant RV motion of the two components, and we need another test to determine if the B star and WR\,2 form a wide binary system or not.

\subsection{Flux ratio measurement from direct imaging}\label{flrat}

\begin{figure}
\includegraphics[width=8cm]{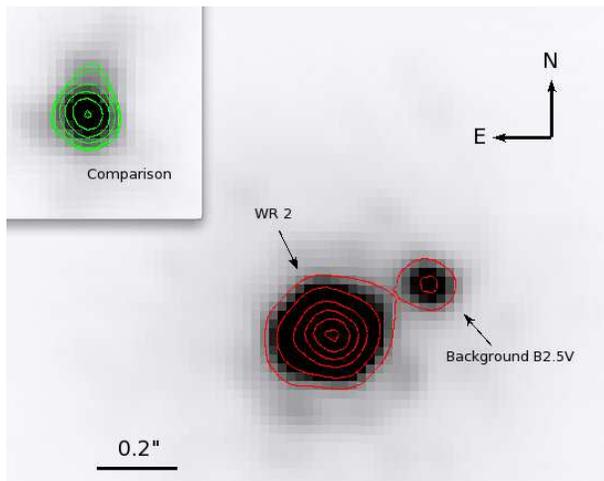}
 \caption{K-band diffraction limited image of WR\,2 obtained with NIRI+Altair at Gemini-North. The contours mark levels between 10 and 15 000 e$^-$/s increasing in a square root scale. The star TYC4017-1829-2,  situated $13.5^{\prime\prime}$ South of WR\,2 is also shown in the left top corner for comparison.}
  \label{PSF}
\end{figure}

Using high-resolution imaging, we successfully detected a faint source at a separation of $0\farcs25 \pm 0\farcs06$ and a position angle of $296\pm5$ degrees with respect to WR\,2 (see Fig.\,\ref{PSF}). No other star is visible down to a radius of $0.05^{\prime\prime}$ (i.e. half the seeing disk FWHM) around WR\,2. Based on the flux ratio and close separation, we suspect that the detected source is the B star whose line we detected in our spectrum . 

Using multiple aperture photometry on the Robo-AO data and subtraction of the primary point spread function, we measure a contrast of 2.4 $\pm$ 0.3 magnitudes in i' (corresponding to a factor 9 difference in flux). NIRI observations reveal a visual companion that is a factor of $\sim$16 fainter (a $\sim$3\,mag difference), located $0.25^{\prime\prime}$ away to the North West from WR\,2 (compatible with the previously mentioned PA). Of particular interest is this NIR flux ratio. Our spectroscopy shows evidence that the B star contributes about 5\% of the flux in the optical, consistent with the 7\% seen here in the NIR (see next section). Despite the extreme luminosity differences, the B star should be almost as bright in the optical as in the NIR if it were at the same distance as WR\,2. As this is clearly not the case, we suggest that the B star is a background star with a slightly higher reddening than that of WR\,2.

\subsection[]{Contribution of the B star}\label{cmfgen}

\begin{figure}
\includegraphics[width=8cm]{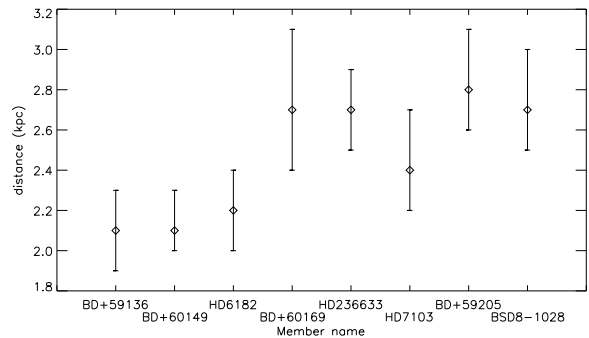}
 \caption{Distance for the Cas OB1 association members using Gaia parallaxes.}
  \label{Gaia}
\end{figure}

WR\,2 is believed to be a member of the Cas OB1 association. Although WR\,2 does not have a direct parallax from Gaia DR2, we can infer its distance using the parallaxes of 8 other association members \citep[noted by][] {Me17}. Two further members were excluded due to their abnormally high and low parallaxes (approximately 0.8mas and 0.04mas respectively) and the remaining member was not observed in Gaia. The proper motions of the members support the conclusion that Cas OB1 is a viable association. Using the distance calculation methods of Rate \& Crowther (in prep), we find the distance for members with Gaia parallaxes (see Figure\,\ref{Gaia}) and calculate a mean of 2.45\,kpc. However, this does not account for the increased error bar size of BD+60 169 and other more distant members. Taking this into consideration, we use a distance of 2.4\,kpc for the association centre and WR\,2.

We have performed a preliminary atmospheric model of the optical spectrum of WR\,2 using CMFGEN \citep{HM98}, which solves the transfer equation in  the co-moving frame subject to statistical and radiative equilibrium, assuming an expanding, spherically-symmetric, homogeneous and static atmosphere, allowing for line blanketing and clumping. CMFGEN does not solve the momentum equation, so the supersonic velocity is parameterized with a classical $\beta$-type law. For this model, we adopted the standard value for the exponent of $\beta$=1. The model yielded an interstellar reddening of $E(B-V)$=0.45 mag for an adopted $R_V$=3.1 extinction law \citep{S79}, corresponding to $E(b-v)$=0.34 mag in the $ubvr$ WR filter system. WR2 has an apparent magnitude of $v$=11.33, so its absolute magnitude is $M_v$=$-$2.1 mag for our adopted distance.

\begin{figure}
\includegraphics[bb=70 205 500 620,width=\columnwidth]{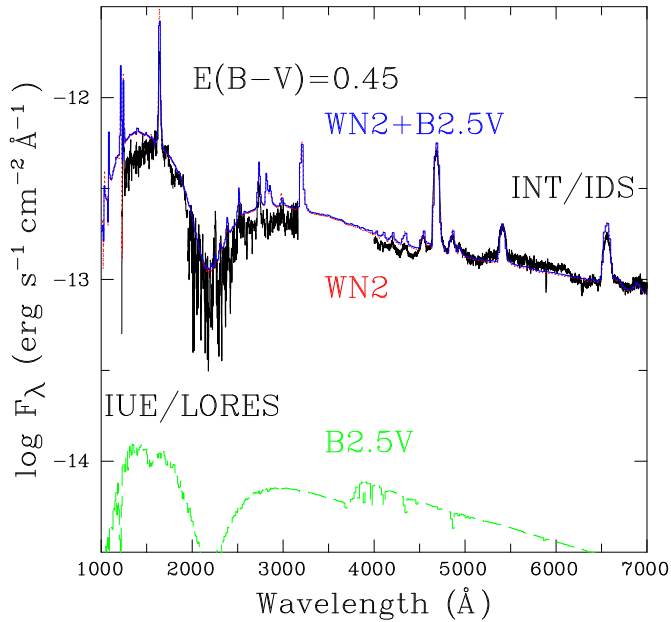}
 \caption{Observed spectral energy distribution of WR\,2 (IUE/LORES, INT/IDS) together with models for the WN2 star (red, dotted line),B2.5\,V (green, dashed lines) and combined system WN2+B2.5\,V (blue, solid line), 
reddened
with $E(B-V)$=0.45 mag.}
  \label{wr2_sed}
\end{figure}

By comparing our mean GMOS spectrum to our WN2 model atmosphere, we estimate an H$\gamma$ equivalent width of $W_{\lambda}$ = 0.30$\pm$0.05\AA\ for the B star. From \citet{MW85}, mid-B dwarfs have H$\gamma$ equivalent widths of 6$\pm$1 \AA, so the contribution of the B star to the spectrum of WR\,2 is approximately 5\% at the wavelength of H$\gamma$. Since $M_v$ = --1.9$\pm$0.5 mag for B2--4\,V stars \citep{MW85}, one would expect a 50\% contribution if the B star was  physically associated with the WN star, or it lay at the same distance. Therefore, assuming the distance to WR\,2 is correct, it is much more likely that the B star is a background line-of-sight companion to WR\,2. Such a conclusion is also supported by the $\sim 2.4$\,mag difference in the i' band between WR\,2 and the B star, giving a distance to the B star of 4 kpc, at a distance of 160 pc from the Galactic mid-plane. While this may seem unlikely, we also note that a B2.5V star has a mass of only $\sim 6 M_\odot$, so it is plausible. In Fig.~\ref{wr2_sed} we present archival ultraviolet (IUE/LORES) and optical (INT/IDS) spectrophotometry of WR\,2 together with our reddened atmospheric model for the WN2 star and a Kurucz ATLAS9 model with parameters representative of a mid-B dwarf ($T_{\rm eff}$ = 19,000\,K, $\log g$ = 4.0) scaled to 5\% of the WN star. Of course, since the B star is a background star, its interstellar reddening would potentially be significantly higher than that of the WN star. 

We conclude that the B-star continuum does not significantly dilute the WR\,2 spectrum and cannot be the principal cause of the emission lines' unusual shape.

\newII{Interestingly, the spectral line amplitudes obtained by our model correspond roughly to those observed, even without the significant contribution of a companion’s continuum. This could be a manifestation of the Baldwin effect \citep{vG01}, i.e. an overall dilution of the WR star's emission lines, as we confirm that WR\,2 has a small radius and a high temperature, potentially leading to a quite strong effect. On the other hand, the shape of the line profile remains unusual, as the models give broad Gaussian-like profiles, in stark contrast with the observed round-shaped lines. We therefore maintain that other phenomena need to be explored to explain fully the WR\,2 spectrum.}

\section[]{Search for evidence of rapid rotation}\label{Rot}

Rapid stellar rotation would cause the geometry of WR\,2 and its wind to depart from spherical symmetry. This can be revealed by a component of intrinsic linear polarization, measurable either in broadband polarimetry or spectropolarimetry. Continuum -- and some high-ionization line -- photons emitted in the inner part of the wind will be scattered by free electrons, and polarized. In a spherically symmetric wind, the individual (vectorial) contributions cancel out. However, if spherical symmetry is broken, as is the case, for example, for a rotationally deformed, peanut-shaped or equatorially flattened wind \citep{Ma99} or any local wind density structure, the net continuum polarization will be non-zero. Emission lines of lower ionization, on the other hand, mostly form through recombination further out in the wind, and therefore are less or even not at all polarized. This leads to the reduction in observed polarization across the emission lines with respect to the adjacent continuum, also called the ``line-effect'' \citep[see][]{Ha98}. This line effect is thus a very sensitive and highly reliable diagnostic for asymmetric winds in massive stars \citep[e.g.][]{St13}. Even in the case of a spherically symmetric {\em rotating} wind, \citet{Ha00} has shown that a line effect can be produced because of the rotational velocity field. For parameters appropriate for the O star $\zeta$ Puppis ($v_{rot}{\rm (equator)}\sim 200\, km\,s^{-1}$), he predicts a line depolarization levels of $\sim$ 0.1\%.

\begin{figure}
\includegraphics[width=8.4cm]{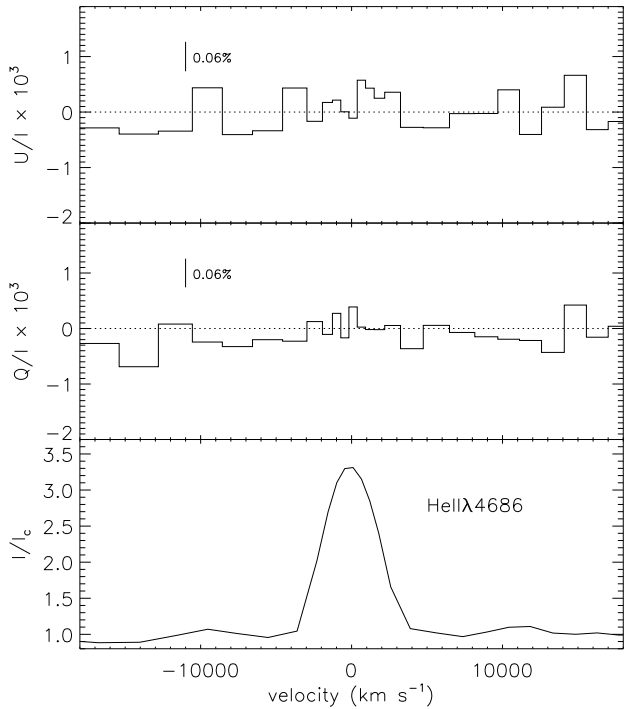}
 \caption{Relative Stokes parameters I/I$_c$ (bottom), Q/I (middle) and U/I (top) for the He\,{\sc ii}$\lambda$4686 line of the hot WN star WR\,2. The 1-$\sigma$ error bars are indicated in the top left corner of the top two panels.}
  \label{figPol}
\end{figure}

\subsection{Observations}

\subsubsection{Linear spectropolarimetry}
A spectrum in polarized light of WR\,2 in the three Stokes parameters I, Q and U was obtained on 2010 August 3 with ESPaDOnS at the CFHT. The spectra were reduced by the pipeline Libre-ESpRIT \citep{Do97}. Each observed spectrum reached a S/N $\sim$ 170 per resolution element (2.6 km\,s$^{-1}$) around 5000\,\AA. Since WR winds have typical turbulent root mean square (rms) velocities of $v_{t}$ $\sim$100~km\,s$^{-1}$ in the radial direction \citep{Le99} (much less in the tangential direction), we chose to bin several resolution elements to obtain a higher S/N. To reach a precision of 0.03\% in the Stokes parameters, we binned the spectra to reach a S/N $\sim$3000 per bin, which had typical sizes between 800 and 1500\,km\,s$^{-1}$ in the continuum. Since the peak of the He\,{\sc ii}$\lambda$4686 line has an intensity $\sim$ 3.5 times higher than the continuum, the spectra need a binning factor that is $\sqrt{3.5}$ times smaller at line peak than in the continuum to reach identical S/N per bin and at the same time produce a reasonable number ($\sim$15) of bins.

One should note that even though ESPaDOnS allows us to obtain reliable polarization relative to the adjacent continuum, this instrument is not designed to obtain accurate measurements of the polarization of the continuum itself \citep[e.g.][]{Mo00}. On the other hand, ESPaDOnS has already been used successfully to detect  line-effects in the spectrum of WR\,1 using the relative Q and U Stokes parameters \citep{St13}.

\subsubsection{Broad-band polarimetry}
We obtained linear polarimetry of WR\,2 using the ``La Belle et la B\^ete'' polarimeter at the Mont M\'egantic Observatory \citep[see][]{Ma95} on the nights of 2010 October 18 and 19. We used three broadband filters (with widths of c. 1000\,\AA\, at $\lambda_c =$ 7500, 5500 and 4300\,\AA), as well as a medium-band filter at 4700\,\AA\, with a bandpass of 100\,\AA. The latter region is dominated by the strong He\,{\sc ii}$\lambda$4686 emission line. We also observed two polarized (HD\,7927 and HD\,19820) and two non-polarized ($\beta$ Cas and HD\,21447) standards in all filters, except the medium-band one where the flux was too low to make this feasible. The former were used to calibrate the polarization angles, while the latter yielded insignificant instrumental polarization at the 0.02\% level. The polarization efficiency ($\sim$80\%) was measured in all four filters using a prism which polarizes the light at 100\% at all optical wavelengths.

The observations of HD\,7927 indicate that the polarization angle does not change significantly with wavelength at the 1$^\circ$ level, which leads us to believe that the angle is also the same in the medium-band filter. On the other hand, unlike HD\,7927, the polarization angle for HD\,19820 deviated from its reported standard value by 50$^\circ$ in all the observed bands, the origin of which remains unexplained.

\subsection{Search for a line effect}\label{specpol}

Fig.\,\ref{figPol} shows the relative Q and U Stokes parameters obtained for the He\,{\sc ii}$\lambda$4686 line, the strongest emission line in the optical band of WR\,2. The mean intensity profile is shown in the bottom panel of the figure. There is no clear signature in the emission lines of a line-effect due to the dilution of polarized continuum light by unpolarized line flux. \citet{Ha98} carried out Monte Carlo simulations to compute polarization spectra of a wind with a latitude dependent density distribution. For the He{\sc ii}$\lambda$5412 line, they found that an equator-to-pole density contrast of 1.5 produces depolarization level of $\sim$ 0.1\%. When significant continuum polarization levels were detected for their sample of stars through the line effect, they report values of $\sim$\,0.6\% , corresponding to an equator-to-pole density ratio of 2-3. Our non-detection, with a 1-$\sigma$ error bar of 0.05\%, corresponds to an equator-to-pole density contrast in their model of $\sim$1.2, leading us to conclude that the wind is highly symmetric with no direct sign of a wind asymmetry that could indicate rapid stellar rotation. The absence of a line effect in our case also indicates the absence of a significant rotational velocity field \citep{Ha00}. This contradicts the assertion that the wind of WR\,2 might have a rotation velocity of 1900\,km\,s$^{-1}$, which is near the break-up velocity and would most likely lead to a very strong wind asymmetry.

\begin{figure}
\includegraphics[width=8.4cm]{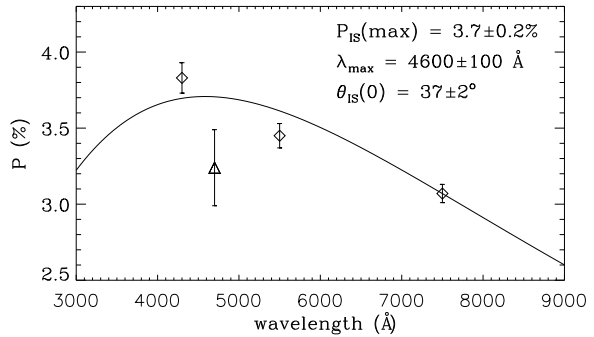}
 \caption{Broad-band polarization observations of WR\,2 (diamonds) superposed on a fitted standard Serkowski law \citep{Se75}. The narrow-band filter observation including the He\,{\sc ii}$\lambda$4686 line is also included (triangle).}
  \label{figPolC}
\end{figure}

\begin{table}
\caption{Linear broad-and medium-band polarimetry of WR\,2}
\begin{tabular}{lccccc}
\hline
$\lambda$ &	(\AA)     &	7500		&	5500		&	4300		&	4700/100\\
\hline
P                &	(\%)	&	3.07(6)	&	3.45(8)	&	3.83(10)	&	3.24(25)\\
$\theta$	 &	(degr)	&    	40(1)		&	35(1)		&	37(1)		&	47(2)\\
\hline
\end{tabular}\label{tabpol}
\end{table}

\subsection{Continuum linear polarization and Serkowski fit}\label{pol}

There is a wavelength dependence of linear polarization seen in essentially all stars that is caused by the ISM, and which follows a Serkowski law \citep{Se75}. We are therefore searching for an intrinsic component of polarization due to electron scattering, which is independent of wavelength in the non-relativistic case. For WR\,2, the final reduced values of the polarization are summarized in Table\,\ref{tabpol}. We fit simultaneously the 3 broad band values of Q = P $\cos{2 \theta}$ and U = P $\sin{2 \theta}$ to Eqs.\,21 \& 22 of \citet{Mo93} for ISM and intrinsic polarization, with 6 free parameters. Only six available data points (Q and U for 3 bands) are not enough to properly fit all the parameters in a meaningful way. Still, after an initial trial, we found that the angle of the ISM polarization ($\theta_{\rm{ISM}}$) does not depend strongly on wavelength, and that in the equation $\theta_{\rm{ISM}}=\theta_{\rm{ISM}}(0)+k/\lambda$, we could assume $k=0$ and reduce the number of parameters to fit down to 5. This first simplification helped, but the fit still remains poorly constrained, as many solutions could satisfy a $\chi^2$-minimization algorithm.

In order to check if the intrinsic polarization of WR\,2 is compatible with a value of zero, as suggested by the linearly polarized spectra discussed in the previous sub-section, we determine what result is obtained when we carried out the fit with an imposed null intrinsic stellar polarization (independent of wavelength). We therefore fitted only 3 parameters: the maximum ISM polarization, $P_{\rm{ISM}}(\rm{max})$, the wavelength at $P_{\rm{ISM}}(\rm{max})$, $\lambda_{\rm{max}}$ and $\theta_{\rm{ISM}}$. The value we obtained for $\lambda_{\rm{max}}$ is $4600\pm100$\,\AA, which is blueward of the average Galactic value that is close to 5500\,\AA. This could simply imply smaller grains on the average towards WR\,2. The measured values of $P$ are plotted in Fig.\,\ref{figPolC} as a function of wavelength. The fitted Serkowski law \citep{Se75}, giving the theoretical curve of the ISM polarization, is over-plotted. It can be seen that the measurements can be reproduced solely with ISM polarization and that there is no need for a non-zero intrinsic polarization of WR\,2. Considering that the reddening for WR\,2 is E$(b-v) =\, $0.4\,mag, the maximum interstellar polarization expected is $P_{\rm{ISM}}(\rm{max})=3.6$\%, according to \cite{Dr03}'s ISM law. This agrees very well with our findings of $P_{\rm{ISM}}(\rm{max})\sim3.5$\%. The narrowband filter containing the strong He\,{\sc ii}$\lambda$4686 emission line deviates only by $\sim2$-$\sigma$ from the theoretical curve at this level. This is compatible with the spectropolarimetric result above, which indicates that there is no significant line depolarization. Nevertheless, higher precision linear pol measurements with better wavelength coverage would be important for more firmly constraining this result.

\begin{figure*}
\includegraphics[]{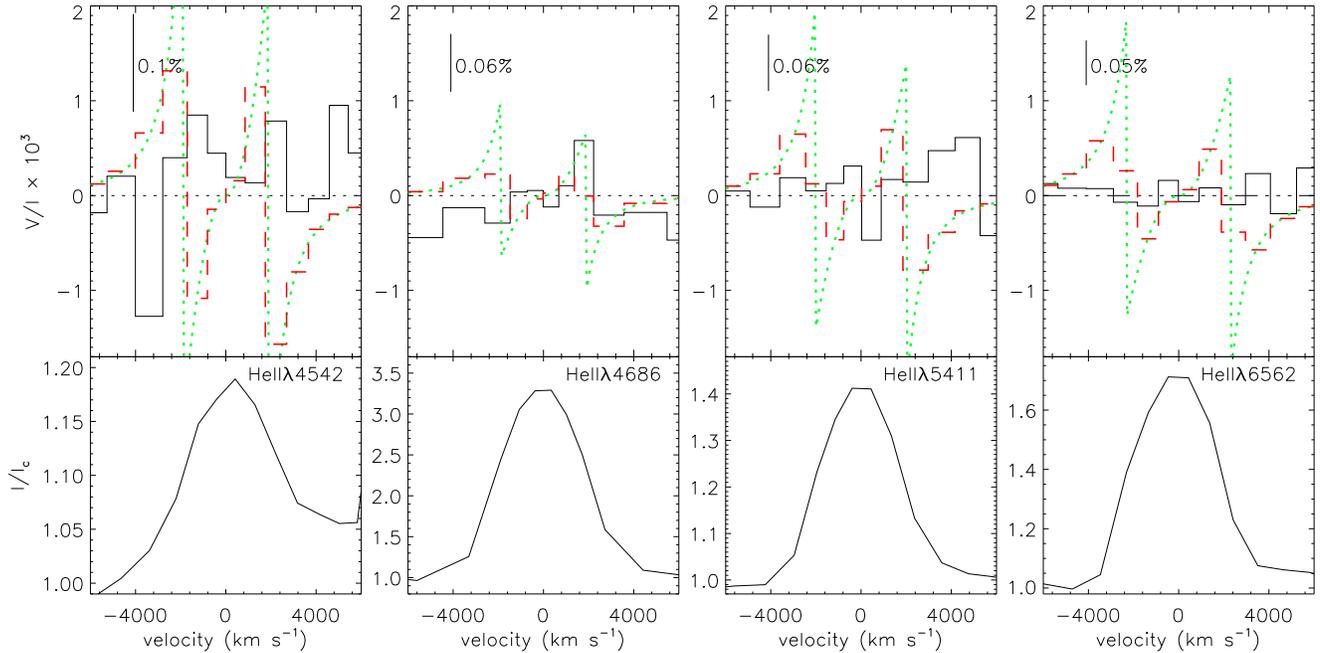}
 \caption{{\it Top panels}~: Observations (black solid line) compared with the split-monopole model (green dotted line) of $V(x)/I(x)$ vs. $\lambda$ of \citet{Ga10}. For a better comparison, the model is also binned into the same velocity sampling as the observations (red dashed line). The $\pm$1-$\sigma$ error bar is indicated in the top left corner. {\it Bottom panels}~: Mean line profiles.}
  \label{figPcirc}
\end{figure*}

\section{Search for evidence of a magnetic field}\label{magfld}

\citet{Sh14} predict that WR\,2 hosts a very strong, organized magnetic field (40\,kG at $R = R_\ast$ and 14\,kG at $R = 1.7 R_\ast$) to enforce the co-rotation required to explain the round-top line profiles in their model. Circular polarimetry can be used to detect magnetic fields via the Zeeman effect in polarization. Using ESPaDOnS at CFHT, we obtained a spectrum in polarized light in the Stokes V parameter on 2014 January 9, as part of the BinaMIcS\footnotemark \footnotetext{BinaMIcS is an international project led by France (PI E. Alecian), which aims to exploit binarity to yield new constraints on the physical processes at work in hot and cool magnetic binary and multiple systems \citep{Al14}.} (Binarity and Magnetic Interactions in various classes of Stars) survey. 

\subsection{Stokes V (circular) spectropolarimetry}\label{specpolV}

Fig.\,\ref{figPcirc} shows in the top panels the V/I signal (plotted with a solid black line) we obtained for 4 emission lines, i.e. He\,{\sc ii}$\lambda\lambda$4542, 4686, 5411 and 6562. The bottom panels show the lines' mean profile. The original S/N was $\sim$ 100 per resolution element (1.8\,km\,s$^{-1}$) around 5000\AA, but we binned the spectra so that they reached a S/N of 2000, 3500, 3500 and 4500 per bin for the four lines, respectively (note that the ESPaDOnS throughput is better towards the red). The observations do not show any clear or coherent signal above 1-$\sigma$ (the uncertainties are plotted as a vertical line in each case). To estimate if our lack of detection of a Stokes V/I signal in those lines is compatible with the values predicted by \citet{Sh14}, we compare our observations to the split-monopole model for a global magnetic field in a WR wind of \citet{Ga10}. For a magnetic pole-on inclination (other inclinations are not considered, but would give qualitatively similar or lower amplitude results) and for optically thick (Sobolev optical depth in the line is much greater than 1) recombination lines ($p = 6$), \citet{Ga10} find (their Eq. 30 \& 32 and in last paragraph of their Section 2.2 as well as their Fig.\,3):\\

$V(x)/I(x) = \left[x/|x|\right] \times \left\{ \begin{array}{ll}
 ^4\!/_3 \, b_1 \, x^2 &\mbox{ for $|x|<1$} \\
 ^{-8}\!/_3 \, b_1 \, |x|^{-3} &\mbox{ for $|x|>1$}
\end{array} \right.$\\

\noindent where $|x|<1$ is for the inner part of the line profile, $|x|>1$ is for the wings and $x = [\lambda - \lambda_0]/\lambda_0$. $x$ is $\pm$ 1 for $\lambda - \lambda_0 \sim \pm $\,HWHM of the line, and\,:\\

$b_1   = 7.7 \times 10^{-4} g_{\rm{eff}}\left(^{\lambda_0}/_{5500 \AA}\right)\frac{B(r)/100 G}{v(r)/100 {\rm{km s}}^{-1}}$ for $r = 1$, \\

\noindent is a factor depending on the magnetic field strength and corresponding to where the line emission peaks.

\begin{figure}
\includegraphics[width=8.4cm]{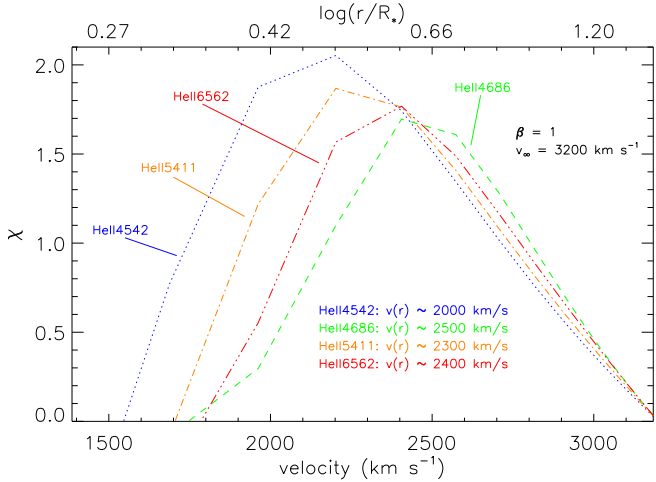}
 \caption{Line formation region emission ($\chi$) calculated with the CMFGEN code for the He\,{\sc ii}$\lambda$4542, 4686, 5411 and 6562 line profiles. The value $\chi$ is plotted as a function of the radius, and also as a function of velocity defined as $v=v_\infty*(1-\frac{R_\ast}{r})^\beta$, where $R_\ast$ is the WR\,2 stellar radius. The value $v(r)$ (given in the plot) is the velocity at which maximum emissivity occurs.}
  \label{emis}
\end{figure}

In \newII{Section\, \ref{cmfgen}}, we described briefly our CMFGEN \citep{HM98} model for WR\,2. From the resulting model we obtained the emitting region of the different lines. 
In Figure\,\ref{emis}, we present our results using $\beta=1$ and $v_\infty=3200$\,km\,s$^{-1}$ (obtained from our CMFGEN model).


We find from Fig.\,\ref{emis}, $v(r ) = 2500$\,km\,s$^{-1}$ for the peak emission of the He\,{\sc ii}$\lambda$4686 line, and thus we adopt that the line emission region is centered at a distance of $R = 4.6 R_\ast (= 3.9 R_\odot)$ from the center of the star. Taking $B(r) = 1900\,G$ at $4.6 R_\ast$ \citep[scaling as $r^{-2}$ from the 40\,kG value at the stellar surface $-$ ][]{Sh14} and $g_{\rm{eff}} \sim 1$, we obtain $b_1 = 0.00050$\,$(0.050\% )$. Using this value, we overplot the modeled $V(x)/I(x)$ vs. $\lambda$ (with a green dotted line) in Fig.\,\ref{figPcirc}  on the data. However, for a fair comparison, we bin the model to the same velocity sampling as the observations (red dashed lines). From this plot we see that the predicted magnetic signal, that sometimes barely reaches 2-$\sigma$, is comparable to the observational uncertainties. The results are therefore inconclusive in the context of the \citet{Sh14} study.

On the other hand, the magnetic field inferred for WR\,2 by \citet{Sh14} assumes a split monopole solar wind-type model \citep{We67} for the field from the stellar surface to the Alfven radius. As discussed by \citet[][Sections 2 and 3]{ud09}, the split monopole model is an idealized case, allowing a quasi-analytic formulation that is not possible with a more realistic dipole field.

The assumption of a split monopole field and its corresponding $r^{-2}$ decay beginning at the stellar surface is conceptually at odds with the picture of a rigidly-rotating magnetosphere in which the wind is confined by the field out to $r=R_{\rm A}=6~R_*$. In fact, the split monopole geometry of \citet{We67} is a simplification which explicitly assumes that the wind dominates the magnetic field at the stellar surface. However, in the picture adopted by \citet{Sh14}, the field must be essentially undistorted by the wind out to the Alfven radius.

Another discordance with previous works in the derivation of \citet{Sh14} is that the wind in a split monopole model corotates up to the Alfven radius. As discussed by \citet{ud09}, most of the angular momentum loss in such a model is actually from the Poynting stresses in the magnetic field, not in the azimuthal velocity of the wind. Thus the wind is actually rotating more slowly than the rigid body value at $R_{\rm A}$. Ultimately, to really force the wind to co-rotate out to 6~$R_*$ would require a magnetic field at the surface of WR\,2 that is substantially stronger than that inferred by \citet{Sh14}.

\subsection{Revised magnetic field predictions}
In this section, we attempt to revise the prediction of WR\,2's magnetic field using more up-to-date assumptions. To calculate the required field, we begin by assuming the simplest possible structure of the unperturbed stellar magnetic field: a dipole aligned with the rotation axis.  We adopt the formalism of \citet{ud02}, more specifically, we used Eq. 4 of \citet{ud09} to compute the wind confinement parameter $\eta(r)$:

\begin{equation}
\eta(r)={\eta_*}{{r^{2-2q}\over{{(1-1/r)^\beta}}}},
\end{equation}

\noindent where $r$ is in units of $R_*$, $q$ is the power-law exponent for the radial decline of the assumed stellar field, e.g. $q=3$ for a pure dipole, $\beta$ is the velocity-law index of the wind, and $\eta_*$ is the wind confinement at the stellar surface. Setting $\eta(R_{\rm A})=1$ and assuming a dipole stellar field, we find that the Alfven radius can be evaluated as the root of the following expression:

\begin{equation}
R_{\rm A}^{4}\, (1-1/R_{\rm A})^\beta={\eta_*}.
\label{eta}
\end{equation}

\noindent \citep[N.B. Eq.\,(\ref{eta}) is just a generalization of Eq. 8 of][to explicitly include the exponent $\beta$.]{ud08} We evaluate Eq.\,(\ref{eta}) for $r=R_A=6~R_*$, which yields $\eta_\ast=1080$ for $\beta=1$. Adopting $R=7~R_*$ for the radius of formation of the bulk of He~{\sc ii}~$\lambda 4686$ ($1~R_*$ outside of the Alfven radius), we can infer the field required by the rigid rotation model that should be reflected in the circular polarization of this line.

We assume that the magnetic field decays as $r^{-3}$. While this is not strictly true outside the Alfven radius (where the field will be ultimately stretched into an approximate split monopole decaying as $r^{-2}$), the proximity of the line formation region and the Alfven radius makes this a reasonable approximation. Moreover, it will provide a conservative upper limit on the field required by the model.

According to Eq. 6 of \citet{ud08}, and considering the radius $R_*=0.85~R_\odot$, mass-loss rate $\dot M=10^{-5.3}~M_\odot$/yr and wind terminal velocity $v_\infty=1800$~km/s \citep{Sh14} of WR\,2, $\eta_\ast=1080$ corresponds to a surface dipole strength of $B_{eq}~=~\sqrt{\eta_*\dot{{\rm M}}v_{\infty }/ R_*^2}=130$\,kG. For comparison, \citet{Sh14} estimated a surface field strength of just 40\,kG (with $\beta=1$). 

Extrapolating these surface fields to $r=7~R_*$ according to the scheme described above, we estimate $B(7~R_*)=380$~G for $\eta_\ast=1080$. Fig.\,\ref{figPcirc} illustrates that this field is well below the noise level of our circular polarization data. 

It is thus clear that even with the much stronger surface fields predicted by our more realistic model of the wind confinement of WR\,2, the more rapid decay of the field strength with distance results in fields in the line formation region that are of comparable strength to that predicted by \citet{Sh14}. Our observations are unfortunately of insufficient quality to confirm or falsify this revised prediction.

However, our revised model still has important implications for the plausibility of the conclusions derived for WR\,2 and the other stars studied by \citet{Sh14}. First, in a manner similar to WR\,2, the surface field strengths required for the other stars must also be substantially stronger for this mechanism to work. For example, to enforce co-rotation of the wind of BAT99-7 (one of the most extreme cases of broad-rounded emission lines in the LMC) out to $16~R_*$ would require a surface field of about 12~MG (mega-gauss!) for $\beta=1$. Such a field corresponds to a surface magnetic flux of $2.5\times 10^8$~G\,R$_\odot^2$, which is vastly larger than those typically inferred in strongly-magnetised upper main sequence stars, white dwarfs and neutrons stars ($\ltsim 1\times 10^6$~G\,R$_\odot^2$). 

Secondly, angular momentum loss due to the strongly magnetized winds of these stars is difficult to reconcile with their inferred rotation speed. In the case of WR\,2, the spindown e-folding timescale ($\tau_{\rm spin}$) according to Eq. 25 of \citet{ud08}, assuming a moment of inertia factor $k=0.1$) is just 15\,000 yr. For BAT99-7, it is only 150 yr!

Finally, the rapid rotation and strong magnetic confinement implied by these models would place these stars in the ``Centrifugal Magnetosphere'' regime of the Rotation-Confinement diagram of \citet{Pe13}. If the axes of their magnetic fields are oblique to their rotation axes \citep[as is almost always the case in magnetic massive stars, e.g.][]{Gr12}, one would expect significant modulation of the magnetospheric emission signatures according to their rotation periods (approximately 0.1~d for WR\,2, 0.4~d for BAT99-7) as observed for other magnetic O stars \citep[e.g. ][]{Ho07}. While the velocity regime of the formation region of the He~{\sc ii}~$\lambda 4686$ emission line and the other lines investigated above lies outside of the magnetospheric region, their zone of formation illustrated in Fig.~\ref{emis} extends significantly into the magnetosphere, and hence these lines should be subject to important rotational modulation. No such modulation is visible on the timescale of minutes to years in the spectra used for the analysis in Section\,\ref{RV}.

\section{Summary and conclusions}\label{Disc}

Our observations and analysis do not support any of the presented scenarios to explain the ``bowler-hat''-shaped broad emission lines as observed in WR\,2, or any implication in four other WR stars in the LMC. Our conclusions are summarized below.

\subsection{Line dilution from a companion}

We have demonstrated that the WR spectrum of WR\,2 is slightly diluted by that of a B-star, but far from enough to explain its spectral lines profile shape. 

The search for RV variations did not show significant Doppler motion of the two components for over 10 years. Both stars, separated by $0.25^{\prime\prime}$, were resolved in our diffraction-limited images, and the B-star is $\sim2.5$\,mag and $\sim3$\,mag fainter than WR\,2 in the optical and NIR, respectively. Considering its brightness, we conclude that it is not bound to WR\,2, but rather lies fortuitously along its line-of-sight. Finally, we have found no sign of another star being either bound or close to the line of sight of WR\,2. 

One should note that the absence of intrinsic continuum linear polarization is unlike (close) binary WR stars which show phase-dependent linear polarization variations and/or variable line depolarization \citep{Ha98}. Also no signs of wind collision have ever been found for this star. This hence reinforces our conclusion that there is no evidence of the presence of a massive companion bound to WR\,2.

\subsection{Rapid rotation}

We have also demonstrated that the wind of WR\,2 is almost certainly spherically symmetric, contradicting the hypothesis that the star is experiencing rapid rotation. 

Our ESPaDOnS linear polarization data reveal no line depolarization in four of the strongest lines in the spectrum. This non-detection of line depolarization is compatible with our broadband linear polarization observations: the observed continuum linear polarization can be accounted for solely by the ISM, so the star is unlikely to be intrinsically polarized. This result implies that WR\,2 is not affected by the von Zeipel effect, meaning that its surface is {\it not} rapidly rotating. We cannot infer that the core is also not in rapid rotation, since it can be decoupled from the surface. Therefore, we cannot conclude with certainty that WR\,2 will never become a LGRB, but since we do not find signs of fast rotation at the surface, there is no reason to consider WR\,2 as a LGRB progenitor. Of course, we cannot exclude the unlikely case that we are viewing the rotation axis pole-on and that therefore, the high degree of symmetry we are finding is only apparent. Indeed, assuming that the linear polarization scales with $\sin^2{i}$ \citep{Br77}, where $i$ is the viewing angle, the expected line depolarization could be as low as 0.1\% (2-$\sigma$ for our observations with ESPaDOnS) for $i=30^\circ$. However, as discussed by \citet{Ku99}, assuming a sample of randomly oriented axes, the probability of observing a star with a viewing angle less than a given $i_o$ is proportional to the integral over solid angle from $i=0$ to $i=i_o$ where $i$ is the polar angle. Hence, the probability $P(i<i_o)$ for observing a star less than a given $i_o$ is $P(i<i_o)=1-\cos i_o$, and for $i_o = 30^\circ$, $P(i<i_o)=13$\%. Moreover, we did explore the possibility to increase the S/N up to 3000 per bin for both $Q$ and $U$ Stokes parameters, which gave fewer than 10 bins in the He\,{\sc ii}$\lambda$4686 line that became, therefore, relatively poorly resolved. Still, even with an improved signal per bin, none of the bins displayed a $Q$ or a $U$ value that exceeded the 2-$\sigma$ error bars, which are in that case 0.06\%. Such a shallow line depolarization is obtained for an inclination of $i\sim7^\circ$, which corresponds to $P(i<7^\circ)=0.6$\%. Finally, the possibility that WR\,2's apparent symmetry is due to the star's inclination can easily be verified definitively by observing the linear polarization of the other ``round-lined'' WR stars first identified in the LMC by \citet{Ru08} (an effort that is beyond the scope of this study).

As an additional note, if WR\,2 was a fast rotator, it would make it a prime candidate for strange mode pulsations (SMPs). Indeed, the most violent SMPs are expected in classical WR stars, and are more likely to occur in fast rotating stars \citep[][]{Gl99}. Recent attempts in detecting cyclic photometric variability with periods ranging from minutes to hours, signs of the presence of SMPs, led to a non-detection \citep{Ch18}.

\subsection{Strong magnetic field}

\citet{Sh14} claim that WR\,2's round-top emission lines require relatively fast rotation along with a strong magnetic field, $B\sim 14$\,kG at $\tau$ = 2/3. While our observations are not compatible with fast rotation, we still need to verify the strength of WR\,2's magnetic field.

Our new Stokes V spectropolarimetry unfortunately cannot be used to test that hypothesis, as the noise is comparable to the signal we are looking for. Future observations with better S/N are needed. On the other hand, we note that many assumptions used by \citet{Sh14} are inconsistent with the more recent efforts in understanding magnetic wind confinement of massive stars. Adopting a more realistic dipole field geometry leads to significantly larger (but still undetectable with the data in hand) predicted surface field strengths that are difficult to reconcile with the inferred and observed properties of WR\,2 (and the other WR stars studied by those same authors).

On a side note, if WR\,2's predicted field was local, rather than global, it would not be easily detected using spectropolarimetry. But since it would generate strong stellar spots, we could expect line profile variability caused by so-called co-rotating interacting regions (CIRs). Perturbations at the stellar surface, such as (magnetic) bright spots, propagate into the wind and, as they are carried around by rotation, create spiral-like density structures \citep{Cr96}. However, only 20\% of WR stars show detectable line profile variability interpretable as CIRs \citep{St09,Ch11} and WR\,2 is not one of them. Our spectroscopic analysis in section\,\ref{RV} corroborate the absence of CIRs. Moreover, a local field is not consistent with the large-scale wind confinement needed by the \citet{Sh14} model.


\subsection{Future work}

We plan in the near future to reassess the physical and wind parameters of WR\,2, and look if an alternative solution with only modest rotation and magnetic field, as suggested by our data, can reasonably reproduce the line profiles, in contrast to the model proposed by \citet{Sh14}.

\section*{Acknowledgments}

ANC gratefully acknowledges support Gemini Observatory, which is operated by the Association of Universities for Research in Astronomy, Inc., under a cooperative agreement with the NSF on behalf of the Gemini partnership: the National Science Foundation (United States), the National Research Council (Canada), CONICYT (Chile), the Australian Research Council (Australia), Minist\'{e}rio da Ci\^{e}ncia, Tecnologia e Inova\c{c}\~{a}o (Brazil) and Ministerio de Ciencia, Tecnolog\'{i}a e Innovaci\'{o}n Productiva (Argentina). AFJM, NSL and GAW acknowledge financial support from the Natural Sciences and Engineering Research Council (NSERC) of Canada. NDR gratefully acknowledges his Centre de Recherche en Astrophysique du Qu\'ebec (CRAQ) Fellowship. ANC and GAW finally gratefully thank A. de la Chevroti\`ere and S. P. Owocki for discussions and input that led to the significant improvement of this manuscript. NDR acknowledges postdoctoral support by the University of Toledo and by the Helen Luedtke Brooks Endowed Professorship. The Robo-AO system was developed by collaborating partner institutions, the California Institute of Technology and the Inter-University Centre for Astronomy and Astrophysics, and with the support of the National Science Foundation under Grant Nos. AST-0906060, AST-0960343 and AST-1207891, the Mt. Cuba Astronomical Foundation and by a gift from Samuel Oschin. Ongoing science operation support of Robo-AO is provided by the California Institute of Technology and the University of Hawai`i. C.B. acknowledges support from the Alfred P. Sloan Foundation. CZ is supported by a Dunlap Fellowship at the Dunlap Institute for Astronomy \& Astrophysics, funded through an endowment established by the Dunlap family and the University of Toronto.
Based on observations obtained at the Gemini Observatory, processed using the Gemini IRAF package. The program ID are GN-2012B-Q-115 and GN-2014B-Q-109.

\label{lastpage}

\end{document}